\documentclass[english,aps,pre,twocolumn,showpacs,superscriptaddress,floatfix]{revtex4-1}
\usepackage[T1]{fontenc}
\usepackage[latin1]{inputenc}
\usepackage{amsmath}
\usepackage{graphicx}
\usepackage{subfig}
\usepackage{amssymb}
\usepackage{dcolumn} % Align table columns on decimal point
\usepackage{bm} % bold math
\usepackage{color}
\usepackage{cleveref}

\makeatletter

\usepackage{amsfonts}
\usepackage{epsfig}
\usepackage{dsfont}

\newcommand{\be}{\begin{equation}}
\newcommand{\bea}{\begin{eqnarray}}
\newcommand{\eea}{\end{eqnarray}}
\newcommand{\ee}{\end{equation}}

\usepackage{babel}
\makeatother

\begin{document}

\title{Role of Intensive and Extensive Variables in a Soup of Firms in Economy to Address Long Run Prices and Aggregate Data}

\author{Ali \surname {Hosseiny}}\email{Al_hosseiny@sbu.ac.ir, mauro.gallegati@univpm.it }
\affiliation{Department of Physics, Shahid Beheshti University, G.C., Evin, Tehran 19839, Iran}
\author{Mauro \surname {Gallegati}}
\affiliation{Department of Economics, Universit\`a Politecnica delle Marche, Italy}

\date{\today}

%\read{jamalitayeb@gmail.com, g\_jafari@sbu.ac.ir, s.vasheghanifarahani@Tafreshu.ac.ir}

%\keywords{Network dynamics, Balance theory, Jammed state, Community, Participation}

\begin{abstract}

 We review the production function and the hypothesis of equilibrium in the neoclassical framework. We notify that in a soup of sectors in economy, while capital and labor resemble extensive variables, wage and rate of return on capital act as intensive variables. As a result, Baumol and Bowen's statement of equal wages is inevitable from the thermodynamics point of view. We try to see how aggregation can be performed concerning the extensive variables in a soup of firms. We provide a toy model for the aggregate production and the labor income as extensive quantities in a neoclassical framework. 

\end{abstract}

%\flushbottom

\maketitle
\thispagestyle{empty}

\section{Introduction} 

The field of complex systems provides a view point to study collective behaviors, attracting attention in different areas of scientific research. It ranges from percolation of disease \cite{{Moreno},{saberia}} to the social sciences \cite{Nekovee}-\cite{Zhou}, complex networks \cite{Newman}-\cite{Baxter}, biology \cite{Peng}-\cite{Mantegna94}, earth sciences \cite{saberib}, economics,  and econophysics \cite{farmer}-\cite{Aoyama2010}, etc. Emergence is one of the key features to distinguish complex systems from regular systems. In thermodynamics for example we start with the kinetic energy and momentum for simple particles of an ideal gas in the micro level. In the aggregate level however to address collective behavior we need to define new parameters such as temperature, pressure, entropy, and free energies which are emerged variables. To address interaction of different systems, in the aggregate level, these variables play a crucial role. In addressing thermal interaction for example, regardless of the micro-structure of systems, temperature indicates direction of the flow of heat and energy. 

In statistical physics, there is good knowledge about important parameters for addressing collective behavior. we can divide state parameters in thermodynamics into intensive and extensive variables. Intensive variables address the equilibrium conditions. 
After relaxation we expect at equilibrium, intensive variables to be the same for interacting systems. Extensive variables however play different roles and define the measures that are needed to be added up in aggregation. Now, one may wonder if in the interaction of sectors in economy we have similar parameters for addressing equilibrium conditions. 

In contrast to physics, in economics, recently the hypothesis of representative agent and actually the hypothesis-of-straightforward-path-from-micro-to-the-macro has been seriously challenged by heterodox schools (see for example \cite{deligatti}-\cite{Schweitzer}). Due to some problems, in economy addressing aggregate behavior is much harder than physics. One problem is that, in contrast to physics, in economics, making new experiments is almost impossible. Physicists usually have been working with table-top experiments and could repeat them frequently. But in economics conducting new experiments is almost impossible. Another difference is that physics usually deals with simple substances, while in economics micro elements are dealt with which creates serious complications. In other words the humans are trying to outguess the market and play with it. This makes things more complicated \cite{arthur}-\cite{chakrabarti}. Another problem with the hard task of aggregation in economics is that in contrast to physics, economics is engaged with systems with fast evolution. Every day new technologies are developed and the relation between different sectors deforms. So, opposite to physics which we can think of long equilibrium, in economics it is the dynamics that we must deal with. 

Economists deal with the problem of evolution by separating long and short run behaviors. Different schools however have their own assumptions \cite{Panico}-\cite{perloff}. In the neoclassical framework it is supposed that economy lives in equilibrium and the Walrasian auctioneer helps different agents to make their mind and find best strategies \cite{Black}-\cite{Samuelsonp}. Some other schools however emphasize the impact of market failure, bounded rationality, and far from equilibrium behavior \cite{stiglitz98}-\cite{simon}.    

In this paper we consider production function and hypotheses concerning its role in addressing long run relations between different sectors of economy in the micro and aggregate level. We discuss similarities of the hypotheses in economy with the hypotheses in thermodynamics.

We should notify that many studies have challenged the hypotheses of the neoclassical framework. There is friction for finding new skills, and the mobility of labor is not easy \cite{{stiglitzunemployment},{aoyama2015}}. In addition, the concept of capital has been debated in the Cambridge Capital Controversy \cite{Robinson1953}-\cite{Solow} (for a review see \cite{cohen}). Even the whole neoclassical synthesis has been criticized \cite{stiglitzrecon}. Recent works emphasize cascades, fragility of the market, and its far from equilibrium behaviour \cite{Kirmann}-\cite{Lorenz}. In this paper however, ignoring all debates, we discuss within the neoclassical framework and claim that within this framework while labor, capital, and production play the role of extensive variables, wage and rate of return on capital resemble intensive variables. 

We revisit Baumol and Bowen's statement on the impact of equal wages in different sectors \cite{Baumol1966}. We claim that their hypothesis is the requirement of the equality of the intensive variables at equilibrium. In other words we revisit the Baumol's cost disease phenomenon \cite{Baumol1989}-\cite{Baumol2012} within a thermodynamics framework. This will provide physicists a  a vision on why expenses for services are growing quickly in developing countries. Baumol's cost disease is a well-known fact in economy. Borrowing some known concepts in thermodynamics we aim to provide an impression on it in physics. We then observe how we can aggregate extensive variables in a soup of firms in a neoclassical framework.

This paper is organized as follows. In Sec. II we review the production function and the Cobb-Douglass model. In Sec. III we discuss the role of intensive and extensive variables in a soup of firms. We revisit Baumol's cost disease phenomenon through a thermodynamics point of view. In Sec. IV we introduce a toy model and aggregate extensive variables in this toy model as a mathematical task. In this toy model we measure some extensive quantities such as production and labor income, and show by keeping the production functions in the micro level and the aggregate level unchanged, the labor share in the aggregate level is dramatically shaped.

%%%%%%%%%%%%%%%%%%%%%%%%%%%%%%%%%%%%%%%%%%%%%%%%%%%%%%%%%%%%%%%%%%%%%%
%%%%%%%%%%%%%%%                                                                                     %%%%%%%%%%%%%%%
%%%%%%%%%%%%%%%             Section II{A Review on Production Function in Micro and Aggregate level
%%%%%%%%%%%%%%%                                                                                     %%%%%%%%%%%%%%%
%%%%%%%%%%%%%%%%%%%%%%%%%%%%%%%%%%%%%%%%%%%%%%%%%%%%%%%%%%%%%%%%%%%%%%

%\section{Intensive Variables, Soup of Firms and Aggregation of Production in a Heterogenous World}\label{secproductionfunction}

%%%%%%%%%%%%%%%%%%%%%%%%%%%%%%%%%%%%%%%%%%%%%%%%%%%%%%%%%%%%%%%%%%%%%%
%%%%%%%%%%%%%%%                                                                                     %%%%%%%%%%%%%%%
%%%%%%%%%%%%%%%             Section II{A Review on Production Function in Micro and Aggregate level
%%%%%%%%%%%%%%%                                                                                     %%%%%%%%%%%%%%%
%%%%%%%%%%%%%%%%%%%%%%%%%%%%%%%%%%%%%%%%%%%%%%%%%%%%%%%%%%%%%%%%%%%%%%

\section{A Review on Production Function in the Micro and Aggregate Level.}\label{secproductionfunction}
Production function is a function that addresses output of a firm in terms of population of hired labors and the value of invested capital
\bea
Q_a=Y_a(T_aL_a,K_a).
\eea
In this equation $Q_a$ is the quantity of good $a$ produced in a firm in sector $a$. The number of labors hired in the firm is indicated by $L_a$, capital is indicated by $K_a$, and productivity is indicated by $T_a$. The function is supposed to have scaling properties as
\bea\label{scaleII}
Y_a(zT_aL_a,zK_a)=zY_a(T_aL_a,K_a).
\eea
Besides, it is supposed to have diminishing return to capital. It means that it is a convex up function of capital.
Now, to choose strategy to hire more labors or invest more capital managers need to look at margins. Hiring more labor and investing more capital are reasonable if increased output compensates for the prices
\bea\label{inequality}
P_a\Delta Y_a\ge \Delta L_a W,\;\;\;\;\;\;P_a\Delta Y_a\ge (R_c+\delta_a)\Delta K_a
\eea
in which $P_a$ stands for the price of units of good $a$, $W$ indicates wage, $R_c$ is the rate of return for capital and $\delta_a$ indicates depreciation rate of capital. In a competitive market we expect
\bea\label{wageproduction}
P_a\frac{\partial Y_a}{\partial L_a}=W\;\;\;\;\;P_a\frac{\partial Y_a}{\partial K_a}=R_c+\delta_a.
\eea

The production function is supposed to address allocation of income. Labors' income in a firm in sector $a$ is $L_aW$ which through Eq. (\ref{wageproduction}) will be
\bea\label{laborshareproduction}
L_aW=L_aP_a\frac{\partial Y_a}{\partial L_a},
\eea
and the income of holders of capital is
\bea\label{capitalshareproduction}
K_a (R_c+\delta_a)=K_aP_a\frac{\partial Y_a}{\partial K_a}.
\eea

Labor share of aggregate income has been almost sustainable in all developed countries for over a century. To justify such stylized fact, Cobb and Douglas in 1928 \cite{cobb} presented their model. They supposed that the whole economy has an aggregate production function as the form
\bea\label{cobbdouglassproduction}
Y=(TL_t)^{\lambda}K_t^{1-\lambda},
\eea
in which $L_t$ stands for the total number of labors in a country and $K_t$ states the aggregate capital. They suggested that by some unknown reasons $\lambda$ should keep a sustainable value over time (around 2/3 for the United States). Plugging aggregate production function from Eq. (\ref{cobbdouglassproduction}) into equations (\ref{laborshareproduction}) and (\ref{capitalshareproduction}) we find a sustainable share of aggregate income for labors.

Since the presentation of the model, the concept of aggregate production function generally and the Cobb-Douglas one especially have been under serious criticism. The model however, at least from a mathematical point of view has been able to justify the historical fact of sustainable share of labor income. In this paper we try to follow a mathematical framework to measure share of aggregate labor income in a toy model.

%%%%%%%%%%%%%%%%%%%%%%%%%%%%%%%%%%%%%%%%%%%%%%%%%%%%%%%%%%%%%%%%%%%%%%
%%%%%%%%%%%%%%%                                                                                     %%%%%%%%%%%%%%%
%%%%%%%%%%%%%%%             Section II Indiference curves and GDP gross rate
%%%%%%%%%%%%%%%                                                                                     %%%%%%%%%%%%%%%
%%%%%%%%%%%%%%%%%%%%%%%%%%%%%%%%%%%%%%%%%%%%%%%%%%%%%%%%%%%%%%%%%%%%%%
\section{Intensive \& Extensive Variables and Aggregation in a Soup of Firms}
%{\bf Intensive Variables, Soup of Firms and Aggregation of Production in a Heterogenous World.}
%Despite its mathematical success to explain sustainable share of labor income, the Cobb-Douglas production function lacks any acclaimed micro-founded explanation. Even if you suppose that production function in micro level for all firms are Cobb-Douglas one, in aggregate level the effective model may not have such form. The concept of aggregate production function itself has been criticized in Cambridge capital controversy. In Cambridge controversy one side leaded by Robinson and Sraffa?? from Cambridge university argued that aggregating capital is meaningless. They mentioned that in production function in one hand we relate long run price to the cost of production and capital. In the other hand capital itself needs price since itself is some sort of goods. Beside many other problems such as interest rate were under debate See for example. The other side was mainly leaded by Samuelson and Solow from MIT, Cambridge, Massachusetts who favor the concept of aggregate production function. See for example?? Since both institute hold a name Cambridge the debate was named Cambridge controversy by Harcourt in ??? who was in British side. For an interesting review over the matter see Cohen and Harcourt.
Aggregation of production is not possible except if we consider Baumol's notification. The first impression was that given the price of production of good $a$ we can indicate wage of labor in sector $a$ through  Eq. (\ref{wageproduction}). In an influential work in 1966 however, Baumol and Bowen \cite{Baumol1966} notified that labors can move amongst sectors and wages should be roughly equal for all sectors. So, they concluded that firms are price taker of wage. They notified that prices should be addressed by Eq. (\ref{wageproduction}). Baumol and Bowen made a conclusion. They stated that sectors with low rate of growth in productivity will face high rate of growth in relative prices. An early consequence is that prices for services will grow faster than manufactured production. The point is that a labor in farming sector has roughly the same wage as a teacher. In 1900 a labor could harvest say 1000 kg of carrot and a teacher could teach 20 students. So, a farmer could pay the price of 50 kg of carrot as tuition fee. Agriculture is a progressive sector and productivity grows in it over time. These days a labor in agriculture sector may harvest 50,000 kg of carrot annually. In education side however we have had stagnation in productivity and a teacher still can teach only 20 students. If we plan to keep tuition fee as the price of 50 kg carrots, then the smart teacher leaves her job. The balance is kept if the price for tuition becomes equal to the price of 2500 kg carrots (of course ignoring capital part in our discussion). So, the conclusion is that tuition fee should grow with a rate higher than growth of prices for the agriculture outputs.

Although we provided an example from the education and agriculture sectors, the conclusion can be extended to all other sectors. Suppose that economy has mainly two classes of sectors, stagnant sectors and progressive sectors. Through the point notified earlier we can conclude that as time goes by prices for stagnant sectors rise comparing to the progressive sectors. This is called "the Baumol's cost disease phenomenon". This prediction of Baumol and his colleagues has been validated in the real world. In recent decades prices of services have grown in all developed countries, to have real numbers and values see \cite{Baumol2012}.

From a thermodynamic point of view Baumol and Bowen's statement is actually the constraint we impose on interacting systems at equilibrium. We suppose that for interacting systems at equilibrium, intensive variables are the same. Let's see what are intensive and extensive variables in a soup of interacting sectors in economy. If you duplicate a sector, then capital and labors are duplicated as well. So, these variables represent the property of extensive variables. It is while wages and rate of return on capital keep their initial values. So, wage and rate of return on capital act as intensive variables. In thermodynamics if two interacting systems have different temperature then we have a flow of heat. Heat flows from the hot system to the cold one. As a result, temperature of hot system decreases and temperature of cold system rises. Heat flows until the two systems reach to the same temperature at equilibrium. At equilibrium if systems are in thermal contact then the intensive variable of temperature becomes the same. If the systems are at mechanical contact then the extended force become the same. Same scenario works in economy.

If in a soup of sectors, wage are not the same between different sectors, then we have flow of labors between sectors. As a result, sectors which offer high wage, are faced with abundant candidate and sectors with low wage will face shortage. Thereby sectors with high wage have the chance to decrease wages and sectors with low wage because of the shortage of labor should increase their wages. So, though there is friction for finding new jobs or earning new skills, in the long run, wages should be balanced in a reasonable manner. Same story goes for the rate of return on capital. If we look at Eq. (\ref{wageproduction}) it is mathematically clear why growth in productivity results in decline in prices. If we consider  productivity in sector $a$ at two different times $t$ and $t^{\prime}$, then scaling properties of Eq. (\ref{scaleII}) guarantees that 
\bea
P_{a_t}T_{a_t}=P_{a_{t^{\prime}}}T_{a_{t^{\prime}}}.
\eea
So, equality of wages clearly results in Baumol's statement.

Equality of wages as a result of the mobility of labors were notified by Baumol and Bowen. Same conclusion was made for the rate of return on capital in \cite{ngai} as a result of the mobility of capital.
%Actually equations (\ref{scaleII}) and (\ref{wageproduction}) guarantee that wage and the rate of return on capital show properties of intensive variables and are not added up when we add up sectors. \\    
 In thermodynamics when we put some systems beside each other and let them have interaction, then theoretically we can find equilibrium conditions. In other words we can find the value of intensive variables and as well the aggregate measure of extensive variables. Now, one may wonder if a similar method can be followed in economy. The answer is that if we accept all hypotheses concerning equilibrium conditions and hypotheses concerning production function we can perform such job.  \\

{\bf - Aggregation of extensive variables in a heterogenous world}\\\\
If we forget all discussions concerning the market failure and as well problems such as "reswitching" problem in the Cambridge capital controversy then we can aggregate extensive values such as production and labor income even in a heterogenous world. To explain how we consider a simple economy as a toy model. Let's suppose we have a simple world and a simple economy with only four products: tractor, power engine, CNC router machine and wheat. Now for production function in each sector we suppose:\\

- Farming Wheat Sector: \\

Forgetting the land, to produce wheat we need wheat itself as seed, power engine and tractor. So, in a general form we can write
\bea
Q_W=Y_W(T_WL_W,N_{WW},N_{WT},N_{WP},N_{WC}).
\eea
in which $Q_W$ is the quantity of wheat produced and $Y_W$ is its production function. $T_W$ an $L_W$ state the level of productivity and the number of labors in our wheat sector of economy. $N_{WW}$ states the measure of wheat itself used as capital in its sector. Actually it is the measure of wheat seeded in the ground. The number of tractors used in this sector is indicated by $N_{WT}$ and the number of power engines is denoted by $N_{WP}$. 

In our notation $W, T, P$, and $C$ sequently indicate "wheat", "tractor", "power engine", and "CNC machine". Though, we included $N_{WC}$ which stands for the number of CNC machines in the agriculture sector, it does not have a direct effect in this sector and we may simply write
\bea
Q_W=Y_W(T_WL_W,N_{WW},N_{WT},N_{WP}),
\eea
which states that the amount of wheat produced in a farm is a function of: labor, amount of wheat seeded, the number of power engines and as well tractors used there. 
\\

- Power engine: \\

To produce power engine machine we need CNC machines and power engine itself. In a general form we can write
\bea
Q_P=Y_P(T_PL_P,N_{PW},N_{PT},N_{PP},N_{PC}),
\eea
in which $N_{PW}$ stands for the amount of wheat in a factory which produces power engine machines and $N_{PT}$ states the number of tractors in that factory and so on. Since, a factory which produces power engine machines has nothing to do with wheat and tractor, in a more specific form we can write
\bea
Q_P=Y_P(T_PL_P,N_{PP},N_{PC}).
\eea
We can go further and write the production function for two other sectors. The point is that in these equations, in production function we have mentioned the quantity of each good as capital.
Actually we have entered quantity of physical capital directly into the production functions for each sector. We have eliminated the parameter of money valued capital in production function to hesitate at least part of problems with measurement of capital. If you let us know the number of labors, the number of tractors, the amount of wheat to seed and the productivity level we can simply predict the amount of wheat which can be harvested in a farm. It does not matter what the price of tractors or other capitals are in our economy. We are using the physical quantities of units of goods used as physical capital in this sector and we can find the physical quantity of production in terms of such physical capitals. 

In a general form in an economy with a variety of goods and services we can write
\bea\label{pfunction}
Q_a=Y_a(T_aL_a,N_{ab}),
\eea
in which $N_{ab}$ indicates the quantity of units of goods $b$ which is used as capital in sector $a$. In this equation $b$ basically ranges for all forms of goods produced in the economy. Now, if we can identify all variables $N_{ab}$, $Q_a$, and the prices of each good then we can aggregate extensive variables in our economy. So, the question is how to identify these parameters in the long run.

If in an economy we have $M$ sectors which produce $M$ forms of intermediate and final goods, to address the aggregate capital in equilibrium we should be able to indicate all values of $N_{ab}$. So, we have $M^2$ variables to be identified. As well we need to deal with the problem of price and indicate the prices of all goods. In other words we have $M$ more variables which we identify as $P_a$. To measure quantities of production or $Q_a$ we should identify $L_a$ and it requires $M$ more values to be addressed. So, to aggregate extensive variables in our economy, in overall we need to identify $M^2+2M$ variables. Now, let's see how many constraints we can impose on the variables.

Producers in sector $a$ can utilize more and more goods $b$ as capital which is indicated by the parameter $N_{ab}$. Increasing this capital is reasonable if and only if its benefit compensates for its price or 
\bea
P_a\Delta Y_a \ge P_b\Delta N_{ab}(R_c+\delta_b).
\eea
This inequality is an extension of the inequality in Eq. (\ref{inequality}) to a heterogeneous world. The right-hand side indicates the price for the raised capital and the left hand side specifies money value of increased production.

In a competitive market at equilibrium we expect
 \bea
P_a\Delta Y_a = P_b\Delta N_{ab}(R_c+\delta_b).
\eea
or
\bea\label{capitalab}
\frac{P_a\partial Y_a} {P_b\partial N_{ab}}=(R_c+\delta_b).
\eea
This equation should hold for all varieties of $a$ and $b$. So, we have $M^2$ constraints. Concerning wages we have $M$ constraints
\bea\label{laborab}
\frac{P_a\partial Y_a} {\partial L_a}=W.
\eea
Given the production functions these constraints identify the price of all goods in terms of wage, rate of return and depreciation rate. In addition to prices these equations identify the ratio of capital per labor in each sector. \\
 
{\bf - A bisector economy as an example}\\

Let's consider a bi-sector economy and impose the above-mentioned constraints and observe the results. We suppose in this bisector economy, production functions for sector $A$ and $B$ are as
\bea\begin{split}
&Y_A=(T_AL_A)^{1-\lambda_{AA}-\lambda_{AB}}N_{AA}^{\lambda_{AA}}N_{AB}^{\lambda_{AB}}
\cr&Y_B=(T_BL_B)^{1-\lambda_{BA}-\lambda_{BB}}N_{BA}^{\lambda_{BA}}N_{BB}^{\lambda_{BB}}
\end{split}\eea
In sector $A$ , the set of constraints in Eq. (\ref{capitalab}) appear as
\bea\begin{split}
&\frac{P_A\partial Y_A}{P_A\partial N_{AA}}=R_c+\delta_A,
\cr&\frac{P_A\partial Y_A}{P_B\partial N_{AB}}=R_c+\delta_B.
\end{split}\eea
Concerning wage we have
\bea\label{hetrolabor}
\frac{P_A\partial Y_A}{\partial L_A}=W.
\eea
We have three similar equations for sector B. If we write these six equations and solve them we find that
\bea\begin{split}\label{priceA}
&P_A=W[T_A^{\lambda_{BB}-\lambda_{AA}\lambda_{BB}-\lambda_{AB}\lambda_{BB}-1+\lambda_{AA}+\lambda_{AB}}
\cr&T_B^{\lambda_{AB}\lambda_{BB}+\lambda_{AB}\lambda_{BA}-\lambda_{AB}}(R_c+\delta)^
{\lambda_{AB}\lambda_{BA}+\lambda{AA}+\lambda{AB}-\lambda_{AA}\lambda_{BB}}
\cr&\lambda_{AA}^{\lambda_{AA}(\lambda_{BB}-1)}
\lambda_{AB}^{\lambda_{AB}(\lambda_{BB}-1)}\lambda_{BA}^{-\lambda_{AB}\lambda_{BA}}
\lambda_{BB}^{-\lambda_{AB}\lambda_{BB}}
\cr&
(1-\lambda_{AA}-\lambda_{AB})^{\lambda_{AA}+\lambda_{AB}+\lambda_{BB}-
\lambda_{AA}\lambda_{BB}-\lambda_{AB}\lambda_{BB}-1}
\cr&(1-\lambda_{BB}-\lambda_{BA})^{\lambda_{AB}\lambda_{BB}+\lambda_{AB}\lambda_{BA}-
\lambda_{AB}}
\cr&]^{1/1-\lambda_{BB}-\lambda_{AA}+\lambda_{AA}\lambda_{BB}-\lambda_{AB}\lambda_{BA}},
\end{split}\eea
in which for simplicity we have supposed that $\delta_A=\delta_B=\delta$. The interesting point is that as we expected improvements in technology in sector $B$ will affect price in sector $A$ through $T_B$ in this equation. For distribution of physical capital we find
\bea\begin{split}
&N_{AA}=L_AW\frac{\lambda_{AA}}{P_A(1-\lambda_{AA}-\lambda_{AB})(R_c+\delta)}
\cr&N_{AB}=L_AW\frac{\lambda_{AB}}{P_B(1-\lambda_{AA}-\lambda_{AB})(R_c+\delta)}
\end{split}\eea
in which $W/P_A$ can be expressed in terms of other variables in Eq. (\ref{priceA}). So, it seems that we can find distribution of capital in each sector. We should notify however that the task is still incomplete. In the RHS we have distribution of labors which so far is unknown. Actually, none of the mentioned equations can address distribution of labors. If we define new variables $n_{ab}=\frac{N_{ab}}{L_a}$, then equations (\ref{capitalab}) and (\ref{laborab}) reduce to
\bea\begin{split}
&P_a\frac{\partial y_a}{\partial n_{ab}}=P_a(R_c+\delta_a),
\cr&P_ay_a=W+\Sigma_nP_bn_{ab}
\end{split}\eea
in which
\bea
y_a(n_{ab})=\frac{1}{L_a}Y_a(L_a,N_{ab})=\\Y_a(1,\frac{N_{ab}}{L_a})=Y_a(1,n_{ab})
\eea
So, as it can be seen $L_a$ is eliminated from all equations and we can only obtain capital per labor in each sector or $n_{ab}=N_{ab}/L_a$. It is because we have supposed that production function have constant return to scale in Eq. (\ref{scaleII}). 

Distribution of labors can be obtained from demand side where consumers are supposed to maximize a utility preference $U(Y_a-\delta_b\Sigma_bN_{ab})$. This optimization in demand side shapes the final formation of economy and actually the portion of each sector there. It then will address distribution of labors. Given the constraint
\bea\label{laborconstraint}
\Sigma_aL_a=L_t,
\eea
through the Lagrange method we can maximize utility
\bea\label{lagrange}\begin{split}
&\frac{\partial U}{\partial L_a}=C
\end{split}\eea
and find distribution of labors. Equations \ref{laborconstraint} and \ref{lagrange} provide $M+1$ more constraints. One of them compensates for the new variable $C$, and $M$ of them are left for our distribution of labors in economy. 

To summarize this section we can say that for an economy with $M$ number of intermediate and final goods, to address the matter of aggregation of extended values we have to identify $M^2+2M$ number of variables. There is $M^2$ variables of $N_{ab}$ which indicates the amount of any good utilized as capital in all sectors.  Prices indicated by $P_a$ as well have $M$ different values for all sectors. The number of labors in each sector as well have $M$ values. Now considering equations (\ref{capitalab}), (\ref{laborab}), (\ref{laborconstraint}) and (\ref{lagrange}) we have $M^2+2M$ constraints to find these numbers and address relation between intensive variables and as well aggregate our extensive quantities. We should mention that one of these variables is dependent to all others and actually identifies rate of return in terms of aggregate capital. \\

%%%%%%%%%%%%%%%%%%%%%%%%%%%%%%%%%%%%%%%%%%%%%%%%%%%%%%%%%%%%%%%%%%%%%%
%%%%%%%%%%%%%%%                                                                                     %%%%%%%%%%%%%%%
%%%%%%%%%%%%%%%           \section{Aggregation of Capital in Simple Examples}
%%%%%%%%%%%%%%%                                                                                     %%%%%%%%%%%%%%%
%%%%%%%%%%%%%%%%%%%%%%%%%%%%%%%%%%%%%%%%%%%%%%%%%%%%%%%%%%%%%%%%%%%%%%

\section{Aggregation of extensive quantities in a toy model }\label{sectionexamples}
%{\bf Aggregation of Labor Share in a Toy Model.}
In the neoclassical school of thought the concept of production function has been widely used in both micro and aggregate level. The Cobb-Douglas production function has been proposed to explain sustainable share of wage income in the aggregate level. Despite its mathematical success, the Cobb-Douglas scenario fails Lucas critique. In other words it lacks any micro based foundation. There have been some efforts in this regards. In \cite{Jones} for example, a Pareto distribution of technology improvement has been suggested as a micro base which shapes the aggregate production function as the Cobb-Douglas one. Such explanations however have not been acclaimed as successful explanations in the mainstream literature.

As stated earlier, in a neoclassical framework, production, capital and labor resemble extensive variables. So, we should be able to aggregate these quantities at least within this framework. In this section we follow the method explained in the previous section to find the aggregate values of the mentioned quantities in a toy models. \\\\
{\bf - A simple economy with no intermediate goods:}\\\\
In our toy model we imagine a very simple economy in which there is no intermediate goods and in each sector, the only goods that is used as capital is itself. As well we make another assumption that production function for each good is a Cobb-Douglas one. In other words we suppose that production function for any sector $a$ is as
\bea
Y_a=T_a^{\lambda_a}L_a^{\lambda_a}N_a^{1-\lambda_a}.
\eea
in which $N_a$ is the quantity of units of good $a$ which has been used as capital to produce itself. Constraint on wages in Eq. (\ref{laborab}) appears as
\bea
P_a\lambda_aT_a^{\lambda_a}(\frac{N_a}{L_a})^{\lambda_a-1}=\frac{\lambda_aP_aY_a}{L_a}=W.
\eea\label{capitall5}
As well, constraint on capital in Eq. (\ref{capitalab}) appears as
\bea
(1-\lambda_a)T_a^{\lambda_a}(\frac{N_a}{L_a})^{-\lambda_a}=\frac{(1-\lambda_a)Y_a}{N_a}=R_c+\delta_a.
\eea
Through these equations, the long run price is obtained as
\bea
P_a=W\frac{1}{\lambda_aT_a^{\lambda_a}}(\frac{N_a}{L_a})^{\lambda_a-1},
\eea
or equivelently
\bea\label{wagehaa}
P_a=W\frac{1}{\lambda_aT_a}(\frac{1-\lambda_a}{R_c+\delta_a})^{\frac{\lambda_a-1.}{\lambda_a}}
\eea
Through these equations we as well find the rate of capital per labor in each sector as
\bea\label{capitallevel5}
\frac{N_a}{L_a}=T_a(\frac{1-\lambda_a}{R_c+\delta_a})^{\frac{1}{\lambda_a}}
\eea
Aggregate capital can be expressed as
\bea
K_t=\Sigma_aP_aN_a=W\Sigma_aL_a\frac{1}{\lambda_aT_a^{\lambda_a}}(\frac{N_a}{L_a})^{\lambda_a}
\eea
or equivelently
\bea\label{capitaltoy}
K_t=W\Sigma_aL_a\frac{1-\lambda_a}{\lambda_a(R_c+\delta_a)}.
\eea
The first point to notice is that Eq. (\ref{capitallevel5}) imposes a constraint on the level of capital in each sector. Once you let us know the level of capital per labor in one sector then through this equation we can let you know the rate of return for capital and through this level we can find the level of physical capital per labor in all other sectors. It can be done since the rate of return on capital plays the role of intensive variable. As long that we accumulate capital through increasing the ratio $N_a/L_a$ in each sector, then the rate of return decreases which causes the aggregate production to rise. The golden level of production in the Solow-Swan growth model is where the rate of return is zero and the whole income for capital holders compensate for depreciation. 

It should be notified that the statement presented here has been under debate in capital controversy. Actually it is suffered from reswitching problem. In reswitching problem we observe that under reduction of interest rate, some technologies may reswitch and thereby the hypothesis of smooth behavior in equations (\ref{capitallevel5}) and (\ref{capitaltoy}) can be challenged which we have ignored in this work. For a review over the matter see \cite{cohen}. 

In these equations everything has been expressed in terms of $N_a/L_a$. None of equations cannot indicate distribution of labors in different sectors and the matter of aggregation for extensive variables is incomplete. To go further we need to know the utility preferences and impose equation (\ref{lagrange}) for all sectors. To complete this task for our case, let's suppose that utility in our simple economy has a simple form as
\bea\label{utilityaggregation}
U=\Pi_a(Y_a-\delta_aN_a)^{\sigma_a}.
\eea
Maximyzing this utility through Eq. (\ref{lagrange}), we reach to
\bea
\sigma_a\frac{U}{(Y_a-\delta_aN_a)}\frac{\lambda_aY_a}{L_a}=C.
\eea
Plugging $Y_a$ from Eq. (\ref{capitallevel5}) leads us to
\bea
L_a=\sigma_a\lambda_a\frac{(R_c+\delta_a)}{R_c+\delta_a\lambda_a}\frac{U}{C}.
\eea
The limited number of labors identifies $C$ and we finally obtain distribution of labors
\bea
L_a=L_t\frac{\lambda_a\sigma_a(R_c+\delta_a)/(R_c+\delta_a\lambda_a)}{\Sigma_b[\lambda_b\sigma_b(R_c+\delta_b)/(R_c+\delta_b\lambda_b)]}.
\eea
So, the aggregate capital is
\bea\label{aggregaatecapitals5}
K_t=\frac{WL_t}{\Sigma_b[\lambda_b\sigma_b(R_c+\delta_a)/(R_c+\delta_b\lambda_b)]}\Sigma_a\frac{\sigma_a(1-\lambda_a)}{(R_c+\delta_a\lambda_a)}.
\eea
The aggregate production is 
\bea\label{aggregateproduction5}
GDP=\Sigma_aP_aY_a=W\Sigma_a \frac{L_a}{\lambda_a}=\;\;\;\;\;\;\;\;\;\;\\ \frac{WL_t}{\Sigma_b[\lambda_b\sigma_b(R_c+\delta_b)/(R_c+\delta_b\lambda_b)]}\Sigma_a\frac{\sigma_a(R_c+\delta_a)}{R_c+\delta_a\lambda_a},
\eea
and the labor aggregate share of income is
\bea
\frac{L_tW}{GDP}=\frac{\Sigma_a\sigma_a(R_c+\delta_a)/(R_c+\delta_a\lambda_a)}{\Sigma_b[\lambda_b\sigma_b(R_c+\delta_b)/(R_c+\delta_b\lambda_b)]}
\eea
As it can be seen extensive variables were aggregated in our toy model. Concerning the aggregate labor share, it is interesting to see that while we fix all production functions in the micro level (the set of $\lambda_a$), still the labor share is dramatically changed by utility preference (the set of $\sigma_a$) in the aggregate level. %So, it seems that a scenario which aims to address sustainable share of labor income in the aggregate and has a micro-base foundation needs to do something with utility preference in that level.

\section{Summary and conclusion}
In this paper we revisited production function and as well Baumol's cost disease phenomenon. We notified that in the language of thermodynamics, labor, capital, and production are extensive variables. This is while wage and rate of return on capital are intensive variables. We tried to aggregate extensive quantities in a neoclassical framework. We introduced a toy model to follow the pattern where we could aggregate extensive variables, namely the aggregate labor income. In our toy model we observed that fixing all production functions in the micro level, still utility preference is a key concept to address extensive variables such as production or labor income in the aggregate level. In this work we considered a unique form of utility preferences for all agents. It would be interesting to check an extension to a heterogeneous utility and its possible impact on the relation between variables.
  
\section*{Acknowledgment}

A. H. would like to acknoledge finantial support from the National Elites Foundation of Iran.
\bibliographystyle{plain}
\bibliography{labor}

{}

\end{document}